\documentstyle[aas2pp4]{article}
\lefthead{He \& Zhang}
\righthead{Pure luminosity evolution for E/S0s}
\input psfig.tex

\begin{document}
\title{Questions on pure luminosity evolution for ellipticals}
\author{Ping He\altaffilmark{2}, and Yuan-Zhong Zhang\altaffilmark{1,2}}
\affil{$^1$ CCAST (World Laboratory), P.O.Box 8730, Beijing 100080,
P.R. China (yzhang@itp.ac.cn}
\affil{$^2$ Institute of Theoretical Physics, Academia Sinica, P.O.Box 
2735, 
Beijing 100080, P.R. China (hemm@itp.ac.cn)}

\begin{abstract}

The explanation for the existence of an excess population of faint blue 
galaxies (FBGs) has been a mystery for nearly two decades, and remains one 
of the grand astronomical issues to date. Existing models cannot explain 
all of the observational data such as galaxy number counts in the optical 
and infrared passbands and the redshift distributions of galaxies. Here, by 
modelling the morphological number counts derived from the Hubble Space 
Telescope, as well as the number counts in optical and infrared passbands, 
and the redshift and color distributions of galaxies obtained from 
ground-based observations, we show that the `FBG problem' cannot be 
resolved if elliptical galaxies are assumed to have formed in an 
instantaneous burst of star formation at high redshift with no subsequent 
star formation events, which is just the conventional scenario for 
formation and evolution of ellipticals. There exist great discrepancies 
between the observed $B-K$ color distribution and the predicted 
distribution for ellipticals by such a pure luminosity evolution (PLE) 
model in the context of the conventional scenario. Neither can the mild 
evolution (i.e., the star formation events have lasted for a longer time 
than those of the instantaneous burst and passive evolution since the 
formation of galaxies) for ellipticals be accepted in the context of PLE 
assumption. The introduction of dust extinction also cannot save the PLE 
models. This conclusion holds for each of the three cosmological models 
under consideration: flat, open and $\Lambda$-dominated. Hence, our 
investigation suggests that PLE assumption for elliptical galaxies is 
questionable, and number evolution may be essential for ellipticals.

\end{abstract}


\section{Introduction}

The number counts of galaxies at deep blue ($B$, $\lambda_{eff}$= 4500 \AA) 
and near-infrared ($K$, $\lambda_{eff}$=2.2 $\mu$m) wavelengths produce 
conflicting results: the $B$-band counts show an excess over the 
no-evolution predictions, and suggest strong luminosity evolution in galaxy 
populations, while the $K$-band counts are well fit by the non-evolutionary 
models. When spectroscopic samples became available, it was found that the 
redshift distributions of galaxies were also consistent with the 
non-evolutionary predictions and strong luminosity evolution would lead to 
a high-$z$ distribution, overestimating the observations. This is the 
common statement of the problem of the excess population of faint blue 
galaxies (FBGs), which remains one of the grand astronomical issues (Koo \& 
Kron 1992; Ellis 1997). To get out of these paradoxes, a number of 
scenarios have been suggested, including: (1) pure luminosity evolution in 
galaxy populations, which increases the distance to which galaxies may be 
seen (Tinsley 1980; Bruzual \& Kron 1980; Koo 1981, 1985; Guiderdoni \& 
Rocca-Volmerange 1990; Gronwall \& Koo 1995; Pozzetti, Bruzual, \& Zamorani 
1996); (2) the choice of a cosmological geometry that maximizes the 
available volume, either by adoption of a low value of the deceleration 
parameter $q_0$ (an open cosmological model) or by introducing a 
cosmological constant $\lambda_0$ (Broadhurst, Ellis \& Shanks 1988; 
Colless et al. 1990; Cowie, Songaila \& Hu 1991; Colless et al. 1993; 
Fukugita et al. 1990); or (3) increasing the number of galaxies at earlier 
times, either by introducing additional populations, which once existed at 
high $z$ but have since disappeared or self-destructed (Broadhurst, Ellis, 
\& Shanks 1988; Cowie 1991; Babul \& Rees 1992; Babul \& Ferguson 1996), or 
by merging, that is by assuming that present-day galaxies were in smaller 
fragments at high redshifts (Rocca-Volmerange \& Guiderdoni 1990; Cowie et 
al. 1991; Guiderdoni \& Rocca-Volmerange 1991; Broadhurst, Ellis \& 
Glazebrook 1992; Carlberg \& Charlot 1992; Kauffmann, Guiderdoni \& White 
1994; Roukema et al. 1997). There are still many uncertainties in the FBG 
problem, and we are in need of more inputs, especially from observations, 
in order to constrain the models.

Great progress has been made recently in observational cosmology through 
the use of Hubble Space Telescope (HST), whose unprecedented imaging 
ability enables galaxies to be segregated morphologically into several wide 
classes (Glazebrook et al. 1995a; Driver et al. 1995; Abraham et al. 1996). 
With morphological data it becomes possible to simplify the modelling of 
FBGs so that each morphological type can be modelled independently, thereby 
reducing the complexity of each individual model (Driver \& Windhorst 
1995).

Following this line of thought, in a previous investigation (He \& Zhang 
1998a, hereafter HZ98) we have modelled the number counts of E/S0 galaxies 
obtained from the Medium Deep Survey (MDS) and the Hubble Deep Field (HDF) 
in the HST I$_{814}$-bandpass ($\lambda_{eff}$=8000\AA), and found that the 
number counts of ellipticals could be well explained by PLE models in any 
cosmological geometry under consideration if ellipticals are assumed to 
have formed at high redshift (say $z_f$=5.0) and thereafter have passively 
evolved (i.e., no further star formation). This is just the traditional 
scenario for the formation and evolution of elliptical galaxies (Eggen, 
Lynden-bell, \& Sandage 1962; Partridge \& Peebles 1967). The models with a 
larger timescale of star formation rate (SFR) which takes the exponential 
decay form cannot reproduce well the number counts, while the models with 
$z_f$ as low as, say $z_f$=2.5, in an open or $\Lambda$-dominated universe 
produce a dramatically high tail or high peak in redshift distributions, 
though they predict the number counts fairly well. In particular, we 
emphasize that the cosmological geometry can not be constrained by such 
number counts for ellipticals, in disagreement with the conclusion of 
Driver et al. (1996), who concluded that flat models dominated by a 
cosmological constant are ruled out from comparison of their E/S0 number 
counts (Driver et al. 1995) with their model predictions. The reader is 
referred to HZ98 for details.

But such a scenario for ellipticals needs to be verified further by 
modelling other Hubble types (i.e., early- and late-type spirals and 
irregulars as separated by HST) and the overall population of galaxies, and 
should be constrained by other observational data such as redshift 
distributions and color distributions obtained from ground-based 
telescopes.

Incorporating other observations and modelling of the other types, we now 
find that the FBG problem cannot be resolved by the conventional scenario 
for elliptical galaxy formation. Though such PLE models can roughly 
reproduce the galaxy number counts in the blue and infrared passbands as 
well as the magnitude limited redshift distribution, there exist great 
discrepancies between the observed $B-K$ color distribution and the 
predicted distribution for ellipticals by such a conventional scenario. The 
inclusion of mild luminosity evolution and dust extinction also cannot save 
the scenario. This conclusion holds in each of the three cosmological 
models under consideration: flat, open and $\Lambda$-dominated. Our 
investigation suggests that number evolution may be essential for 
ellipticals.

Number evolution models have been widely considered to account for the 
paradoxes concerning the FBG problem. One of the considerations on number 
evolution is merging between galaxies, which is a natural feature of 
hierarchal theory for the growth of structure by gravity. Notice that a new 
type of models, i.e., semi-analytic models have appeared recently (Cole 
1991; White \& Frenk 1991; Lacey et al. 1993; Kauffmann, White \& 
Guiderdoni 1993; Kauffmann et al. 1994; Cole et al. 1994; Baugh, Cole \& 
Frenk 1996), which are well physically motivated by cosmological theory, 
and may eventually provide the greatest understanding of galaxy formation 
and evolution (Gardner 1998). The traditional galaxy number count models 
(e.g., Yoshii \& Takahara 1988; Guiderdoni \& Rocca-Volmerange 1990; 
Gardner 1998), however, are still powerful tools to explore the formation 
and evolutionary history of galaxies, and can be treated as complements to 
the semi-analytic techniques. We will make use of the traditional approach 
throughout this work.

Finally, we construct a set of simple number-luminosity evolution (NLE) 
models to explain the number counts and color-selected redshift 
distributions in the $B$-band for elliptical galaxies. We argue that the 
number evolution is necessary for reconciling the conflicts 
above-mentioned. In particular, if the cosmic geometry is open or 
$\Lambda$-dominated, our work supports the idea that ellipticals formed by 
mergers of spiral galaxies, which is compatible with the prediction of 
hierarchical theory for formation and evolution of galaxies.

In section 2 we briefly review the procedure for the construction of the 
models. The results of the models are shown in Section 3 for comparison 
with the observational data, and a phenomenological number evolution model 
is presented in Section 4. We will give the summary and conclusions in 
Section 5.

\section{Construction of models}

We employ the latest version of the galactic spectral synthesis models of 
Bruzual \& Charlot (1997), with solar metallicity, to compute the 
evolutionary spectral energy distributions (SEDs) for the galaxies. In the 
BC code, model galaxies are characterized by the initial mass function 
(IMF) and the SFR. Throughout this work, we assume the standard Salpeter 
IMF (1955) for the other Hubble types than early-type galaxies, while a 
Scalo IMF (1986) for ellipticals. The Scalo IMF is more suitable for 
ellipticals than the Salpeter IMF in that a less steep IMF at the high-mass 
end such as the Salpeter one will lead to more massive stars existing at 
early times, rendering UV fluxes are so strong that more galaxies can be 
detected at high $z$, and hence, even with dust extinction involved, the 
models will overpredict the number counts for ellipticals between 
$I_{814}\sim$ 19 and 21 (cf. HZ98). The SFR, $\psi(t)$, is chosen as an 
exponential decay form with respect to time $t$, i.e., $\psi(t)\sim 
exp(-t/\tau_e)$, where $\tau_e$, measured in Gyr, is the timescale 
characterizing this form of the SFR. $\tau_e$ is treated as a free 
parameter to be adjusted to reproduce the local photometric and 
spectroscopic properties, such as present-day colors and spectra for 
galaxies. For comparison, we list the local observed colors of galaxies in 
Table 1. We have also considered to some extent the effect of dust 
extinction, following Wang's (1991) prescription and similar to that of 
HZ98. The extinction magnitudes in the $B$, $I$, and $K$ passbands for 
galaxies of a typical luminosity $L^*$ are listed in Table 2. By adopting 
both the Scalo IMF (for ellipticals) and the dust extinction, the UV flux 
at early times can be greatly reduced so as to avoid the detection of a 
large number of galaxies at high-$z$, which are not observed in current 
deep surveys.

Predictions for the color distribution of galaxies in a given range of 
apparent magnitude $[m_{\lambda1},m_{\lambda2}]$ can be computed by 
integrating the following equation over $m_{\lambda}$ and $z$ ($z \leq 
z_{max})$, as: 
$$N(c)=\int\limits_{c<c(z)<c+dc}\int\limits_{m_{\lambda1}}^{m_{\lambda2}}d^
2N(m_{\lambda},z)d{m_{\lambda}}dz,\eqno(1)$$ where 
$z_{max}=min(z_{f},z_{L})$, with $z_f$ being the formation redshift for 
galaxies, and $z_L$ being the redshift above which the Lyman continuum 
break enters the $\lambda$-filter. $c(z)$ is the color-redshift relation 
for galaxies. In order to compute the differential color distribution 
$dN(c)$, which is a function of color $c$, the integration over $z$ is 
restricted by the inequality $c<c(z)<c+dc$, such as the color $c(z)$ is in 
the relevant color bin $dc$ in the color range [$c$ , $c+dc$]. In the above 
integral, $d^2N(m_{\lambda},z)$ refers to the differential number counts 
for galaxies in the intervals [$m_{\lambda}$ , $m_{\lambda}+dm_{\lambda}$] 
and [$z$, $z+dz$]. Its explicit expression, as well as the formulae which 
we need in this work to compute the number counts and redshift 
distributions, can be found in, e.g., Guiderdoni \& Rocca-Volmerange (1990) 
or HZ98. To make the predictions more realistic, we smooth the differential 
color distribution $dN(c)$ in the colour interval $c\sim c+dc$ by a 
Gaussian profile with $\sigma=0.20$ mag before deriving the integrated 
distribution, which is slightly different from the prescription of Pozzetti 
et al. (1996), while similar to that of Gardner (1998). This procedure is 
expected to mimic the observational errors in the colours as well as the 
intrinsic dispersion in the colours of galaxies of the same Hubble type.

Following the prescription of HZ98, we adopt three representative 
cosmological models in this work: 1) flat, $\Omega_0=1.0$, $\lambda_0=0$, 
and $h=0.5$ ($H_0$=100 $h$ km s$^{-1}$ Mpc$^{-1}$) (hereafter, Scenario A); 
2) open, $\Omega_0=0.1$, $\lambda_0=0$, and $h=0.5$ (Scenario B); and 3) 
flat and $\Lambda$-dominated, $\Omega_0=0.2$, $\lambda_0=0.8$, and $h=0.6$ 
(Scenario C). We assume a formation redshift $z_f=5.0$ for all types of 
galaxies in all three cosmological models.

\section{Results}

\subsection{Luminosity functions}

The luminosity functions (LFs) of galaxies, which is the distribution law 
of the number density of galaxies against their absolute luminosities, is 
an important ingredient for modelling. However, the LFs for present-day 
galaxy populations are not well-determined by local surveys, and the 
universality of LFs are doubtable. Hence we treat the parameters of LFs as 
free to be adjusted to give the best fit of the morphological number 
counts. The LFs take the Schechter form (Schechter 1976) for present-day 
galaxies, and we list the parameters ($\alpha$, $\phi^*$, and $M^*_B$) of 
these ($B$-band) LFs in Table 3. LFs in other pass-bands can be obtained by 
shifting corresponding present-day colors. The model LFs are also shown in 
Figure 1 to compare with the observed ones. It can be seen from Figure 1 
that, in both the $B$- and the $K$-bands, the faint-end slopes of model LFs 
are steeper than those of the observed ones whatever the cosmological 
models will be. The reason for such adoptions will be given in Section 
3.3.1.

Besides, the PLE models alone cannot reproduce the steep slope of the 
number counts at faint magnitudes in the I$_{814}$-band for late-type 
spirals/irregulars in a flat cosmological model (Scenario A), so we enhance 
the characteristic number density $\phi^*$ of the LF for the Sdm galaxies 
with respect to look-back time ${\delta}t$ as 
$\phi^*({\delta}t)={\phi_0^*}(\beta{{\delta}t\over{t_0}})$ ($\beta >0$), 
where the subscript `0' refers to present-day values.

\subsection{Star formation rates}

The parameter $\tau_e$ of the SFR for galaxies is adjusted to reproduce the 
local photometric and spectroscopic properties of galaxies, such as local 
colors and spectra (cf. Fukugita, Shimasaku, \& Ikhikawa 1995; Yoshii \& 
Takahara 1988). We summarize the values for this parameter for all the 
Hubble types in column 2 of Table 3. In particular, the case of $\tau_e=0$ 
for ellipticals (Scenario B) represents {\em purely} passive evolution and 
a constant SFR for the Sdm galaxies is denoted by $\tau_e=\infty$. It can 
be seen from Table 3 that the modelled colors are close to the 
corresponding observed ones (see Table 1). Considering that there are 
uncertainties of $0.1-0.2$mag in colors, the values of $\tau_e$ of our 
models are acceptable.

In the following, we will examine whether the models based on these LF and 
SFR parameters can be accepted by comparing the modelled galaxy number 
counts, redshift and color distributions with those observed.

\subsection{Number counts}

\subsubsection{Morphological number counts}

Corresponding to the morphological classification of galaxies by HST, we 
incorporate Sab and Sbc galaxies into early-type spirals, and Scd and Sdm 
galaxies into late-type spirals/irregulars. With $\beta=1.5$ for Scenario A 
(see above for the reason of the introduction of $\beta$), we can see from 
Figure 2 that the models can reproduce well the number counts in the 
$I_{814}$-band for all three galaxy types as well as for the overall 
population in any cosmological model under consideration. We see that, in 
order to reproduce the number counts for early-type spirals and late-type 
spirals/irregulars at faint magnitudes with the least additional 
assumptions other than PLE, it is necessary to adopt steep faint-end slopes 
of LFs for those galaxy populations.

\subsubsection{Number counts in blue and near-infrared bands}

From Figure 3 we can see that the models can reproduce better the number 
counts only at bright magnitudes in both the $B$ ($B<21$) and the $K$ 
($K<17$) bands, indicating that the normalization (involving both 
$\phi^*_0$ and $L^*$, cf. Ellis 1997) is proper by the adoptions of 
LFs{\footnote {Due to large uncertainties in determining the present-day 
LFs, the normalization of LFs is usually chosen to scale the model 
predictions to the observed number counts at a fainter magnitude, say $B 
\sim 19.0$, rather than at the brightest end, see Pozzetti et al. 1996 for 
a comprehensive review about this.}}. The discrepancies between the model 
predictions and the observations exist at the faintest magnitudes in both 
$B$- and $K$- bands. It seems that there are not sufficient blue galaxies 
to reproduce the steep slope at the faintest magnitudes in the $B$-band, 
whereas in the $K$-band, it seems that too many red galaxies are predicted 
over the observations. These red objects are ellipticals, which dominate at 
most of the $K$ magnitudes in contrast with the case in the $B$-band. 
However, given the uncertainties in faint galaxy number counts due to 
either small sample sizes or galaxy clustering, we can still consider the 
PLE predictions to be in rough agreement with the observations, but we 
should emphasize that the agreement is achieved by the adoption of steep 
slopes for LFs of late-type galaxies. If the slopes are as low as $\alpha 
\sim -1$, the discrepancies are expected to be large, especially for the 
flat Einstein-de Sitter universe (Scenario A).

\subsection{Redshift distribution for $K<20$}

The spectroscopic studies with the LRIS spectrograph on the Keck Telescope 
of two of the Hawaii deep survey fields SSA 13 and SSA 22 (Cowie et al. 
1996) present one of the largest and deepest redshift samples, and provide 
a powerful tool for understanding the evolution of galaxies. Compared with 
optical light, the absolute $K$-band magnitude is mostly contributed to by 
near-solar-mass stars in which make up the bulk of galaxies. It is 
therefore a good tracer of the baryonic mass and particularly suitable for 
the study of old stellar populations. Furthermore, the SEDs at low 
redshifts at near-infrared wavelengths for different morphological classes 
of galaxies are very similar, and hence the morphological mix will be 
insensitive to redshift if there is no evolution, and any evolutionary 
signature will be more clearly seen (Glazebrook 1997).

From the Cowie et al. (1996) sample, we have derived the redshift 
distribution limited at $K<20$, with the total number being 207, and the 
completeness being 100\% for $K<18$ and 72\% for $18<K<20$. We compare 
predictions of our models with the observed data in Figure 4. It can be 
seen that our models predict more high-$z$ objects than the observation in 
the three Scenarios, and most of them are ellipticals. However, given that 
the redshift data for $K>18$ is only 72\% complete, one might suppose that 
much of the high-$z$ discrepancy here is due to incompleteness (see the 
comment on the incompleteness by Pozzetti et al. 1996), we argue that the 
PLE models cannot be ruled out from comparison with the observed data.

\subsection{$B-K$ color distribution for $17<K<20$}

Since the redshift distribution from the Cowie et al. (1996) sample is 
complete to only $\sim 70$\% for fainter magnitudes in $18<K<20$, the 
failure existing in the PLE models can be exposed by using the $B-K$ color 
distribution for galaxies from $17<K<20$ in the same Hawaii sample, whose 
completeness in the $B-K$ color statistics limited in $17<K<20$ is 100\%. 
From Figure 5, we see that none of the models can satisfactorily reproduce 
the observations. In any case, the colors of early- and late-type spirals 
and irregulars will not be redder than $B-K\sim5.5$ according to our 
spectrophotometric models, and ellipticals should be responsible for the 
objects at the red-end of the distribution. But the models underpredict the 
data between $B-K=5.0-8.0$ and produce extremely red humps beyond 
$B-K=8.0$. The great discrepancies between the observations and the model 
predictions indicate that the models need substantial revising. It should 
be pointed out that the predicted color distributions are not sensitive to 
the specific adoptions of LFs, but sensitive to the star formation rate of 
galaxies.

It can be seen that the $B-K$ color distribution for galaxies can 
efficiently reveal the defects of the PLE models based on the assumption of 
{\em purely} passive evolution of ellipticals.

\subsection{Mild evolution in luminosity for ellipticals}

From the above analysis of the $B-K$ color distribution for galaxies, we 
find that far more extremely red ($B-K > 8$) ellipticals than observed are 
predicted by our PLE models in the context of the traditional scenario for 
formation and evolution of ellipticals, and these models are not 
acceptable. For the sake of comparison, we reconsider the modelling for 
ellipticals with the assumption of mild luminosity evolution. {\it Mild 
evolution} here means that the star formation events have lasted for a 
longer time than those of the instantaneous burst and passive evolution 
since the formation of galaxies. Explicitly speaking, the SFR timescale 
$\tau_e$ for mild evolution should be slightly larger than that for passive 
evolution. By assuming $\tau_e$=1.0 Gyr (the modelled $B-K$ colors are 
4.07, 4.18, and 4.18 for Scenario A, B, and C, respectively), we can see 
from Figure 6a that such PLE models as mild luminosity evolution 
overpredict the number counts at faint magnitudes in the $I_{814}$-band for 
Scenarios B and C, and hence these mild evolution models are also not 
acceptable. However, the model prediction for Scenario A appears to fit the 
data reasonably well.

From the previous analysis of color distribution in Section 3.5, we have 
found that ellipticals can be discriminated from the other Hubble types by 
$B-K$ color, so that no objects except ellipticals can be redder than 
$B-K=5.5$. It is worthwhile mentioning that the significance of 
color-selected sub-sample for investigating the evolution of ellipticals 
has been first realized by Kauffmann, Charlot, \& White (1996), but their 
scheme of selecting ellipticals by means of color is slightly different 
from ours, in which we pick out ellipticals by $B-K$ color solely rather 
than the color-redshift relation. Applying this color-selection criterion 
to the current case, we derive from the Cowie et al. (1996) sample the 
redshift distribution for galaxies limited to $22.5<b_j<24.0$ and subject 
to the condition $B-K>5.5$. It can be seen from Figure 6b that 
observationally, ellipticals are completely absent at high-$z$ beyond 
$z=0.8$, consistent with a recent report by Zepf (1997) from the analysis 
of deep optical and infrared images derived from HST. However, our Scenario 
A model underpredicts the objects in the intermediate redshift interval 
($0.3<z<0.8$), and overpredicts the high-$z$ distribution, extending to 
redshifts as high as 1.4, completely in disagreement with the observations. 
Hence, we arrive at the conclusion that the PLE models with mild evolution 
cannot be accepted either in any cosmological model under consideration.

\section{Number-luminosity evolution for ellipticals}

As demonstrated above, PLE models solely are not the appropriate 
simulations for formation and evolution of ellipticals. In any realistic 
evolutionary models for galaxies, however, luminosity evolution must be 
considered due to the aging of stellar populations, which leads to the 
continuous change of photometric and spectroscopic properties of galaxies. 
For simplicity, however, we use the same galaxy evolution models with mild 
luminosity evolution as in the above PLE models (Section 3.6) to compute 
the $K$- and $e$-corrections for galaxies, without considering more 
complicated star formation scenarios (e.g., Col\'in, Schramm, \& Peimbert 
1994; Fritze-v.Alvensleben \& Gerhard 1994), or various complicated 
physical processes concerning the formation of galaxies (cf. Roukema et al. 
1997).

Without considering the underlying physical mechanism, we 
phenomenologically formulate the number evolution for ellipticals by the 
expression $\phi^*(z)=\phi^*_0 f_{\phi}(z)$, where $f_{\phi}(z)$ represents 
the following function:
\setcounter{equation}{1}
\begin{eqnarray}
  f_{\phi}(z)=\left\{
	\begin{array}{ll}
	   (1+z)^{-Q_{n1}}, & z \leq z_{t};  \\
	         &            \\
	   (1+z_{t})^{-Q_{n1}} ({{z-z_f} \over {z_t-z_f}})^{Q_{n2}}, & z_t 
< z \leq z_f,
	\end{array}
       \right .
\end{eqnarray}
and for consistency, we assume the characteristic mass $M^*$ of mass 
function (MF) for galaxies decreasing with respect to redshift $z$ as 
$M^*(z)=M^*_0 f_{M}(z)$, where $f_{M}(z)$ represents the following 
function: $$f_{M}(z)=(1+z)^{-Q_m}, \eqno(3)$$ and for simplicity, we assume 
the faint-end slope $\alpha$ of LF does not change with $z$, and the 
evolution of LF maintains the property of self-similarity. Hence the number 
evolution model is characterized by the parameters $Q_{n1}$, $Q_{n2}$, 
$Q_m$ and the redshift $z_t$. The MF can be translated into LF by the 
present-day mass/luminosity ratio of galaxies, and the evolution of LF 
exactly follows the same formulae as Eq. [2] and [3]. We expect such a 
simple treatment can to the most extent reflect the basic features of the 
evolutionary history for ellipticals without getting bogged down into too 
many technical details. Our NLE model presented here, if $z_t=z_f$, is 
similar to the $M^*$--$\phi^*$ model designed by Guiderdoni \& 
Rocca-Volmerange (1991). Obviously, the NLE model degenerates to PLE model 
when $z_t=z_f$ and $Q_{n1}=Q_m=0$.

We have tentatively determined a set of these parameters ($z_t$, $Q_{n1}$, 
$Q_{n2}$, and $Q_m$) by trial and error, whose values are listed in Table 
4. We show the predictions of the NLE models and compare with the observed 
number counts for ellipticals in Figure 7a. We see that, in contrast with 
the predictions of PLE models with mild luminosity evolution presented in 
Section 3.6, the predictions of the NLE models are satisfactory; in 
particular, they can reproduce well the faint-end counts as well as the 
flattening in any world model under consideration. We have also presented 
in Figure 7b the predictions of the $z$-distributions to compare with the 
same color-selected sample as in Section 3.6. It can be seen that our model 
predictions are in gross agreement with the color-selected sample for 
Scenario B and C, and do not show significant high-$z$ tails. The NLE 
prediction for Scenario A is better than the case of PLE, but it still 
overestimates the high-$z$ distributions. This discrepancy between the 
model prediction and observed data should be regarded not as the rejection 
of the flat cosmological model ($\Omega_{0}=1$) but, we believe, as the 
indication that the model needs further elaborating, since we have not 
considered, for example, the evolution of metallicity with respect to $z$, 
and the $(1+z)^4$ dependence of the surface brightness with respect to $z$ 
for the sake of simplicity (cf. Yoshii \& Peterson 1995). As addressed by 
Pozzetti et al. (1996), the magnitude of this effect not only is a function 
of various intrinsic parameters of galaxies, but also depends on the data 
reduction procedure. This effect, if considered in our models, could be 
expected to further reduce the number of ellipticals at high $z$.

In Figure 8 we show the predicted $B-K$ color distributions with the mild 
luminosity evolution and the number evolution which follows the expression 
of Eq. [2] and [3]. It seems that the NLE models do not reproduce 
satisfactorily the $B-K$ color distribution, but we can see that the 
red-end humps have disappeared, which to some extent indicates the success 
of the models. As for the discrepancies between the predictions and the 
observation, we argue that they might reflect that the spectrophotometric 
models need further elaborating, e.g. fine-turning the SFR timescale 
$\tau_e$ for ellipticals or dividing ellipticals into several sub-classes 
with different $\tau_e$ specified for each of them. We will consider these 
possibilities elsewhere (cf. He \& Zhang, 1998b).

\section{Summary and Conclusions}

The assumption of pure luminosity evolution is no doubt the starting point 
for investigating the formation and evolution of galaxies. Traditionally, 
ellipticals are believed to evolve passively after they formed in an 
instantaneous burst of star formation, and indeed the models based on such 
an assumption can reproduce fairly well the number counts for ellipticals 
(HZ98), but in this work we find that such PLE models completely fail to 
account for the $B-K$ color distribution limited in the faint $K$ magnitude 
bin of $17<K<20$. On the other hand, the models with mild luminosity 
evolution in the context of PLE assumption will overestimate the faint-end 
elliptical number counts in an open or $\Lambda$-dominated cosmological 
models (Scenario B and C), or overpredict the high-$z$ objects in the flat 
universe (Scenario A), significantly in disagreement with the 
color-selected redshift distribution (see Figure 6), and hence also cannot 
be accepted. It seems that the PLE models for ellipticals, whether the 
luminosity evolution is purely passive, as assumed by traditional scenario, 
or mild, cannot reconcile simultaneously the number counts and color 
distribution for ellipticals. Therefore, these paradoxes indicate that the 
evolution within the population of elliptical galaxies may be more 
complicated than that expected by the PLE assumption.

These conflicts exist in all three cosmological models considered here, 
indicating that changing cosmological geometry is not a `graceful exit'. 
Furthermore, dust extinctions will both shift the red-end humps even redder 
and decrease the number of ellipticals in the $B-K$ color distribution, and 
hence also cannot save the traditional scenario for ellipticals. We are 
obliged to abandon the pure luminosity evolution scenario for ellipticals 
when facing these conflicts.

We have not restricted ourselves to any `observed' LFs due to the various 
uncertainties in determining them locally. Models based on such `observed' 
LFs can not even reproduce the morphological number counts. Instead, we 
treat the LF parameters $\alpha$, $\phi^*$, and $M^*$ as free parameters to 
be adjusted to reproduce the morphological number counts. Most importantly, 
our results, e.g., the predicted $B-K$ color distributions, do not 
critically depend on specific adoptions of LFs. Our conclusions are robust 
unless the observational data are plagued with great uncertainties or 
biases, such as errors in morphological classification for galaxies, 
selection effects, or even unrealized systematic errors.

If the assumption of PLE has to be abandoned, then number evolution for 
ellipticals should be a rational inference. As mentioned in Introduction, 
an additional galaxy population is usually considered as a plausible 
proposal to account for the excess faint galaxies observed at optical 
bands, which once existed at early epochs but disappeared or 
self-destructed subsequently, and hence cannot be detected by local 
surveys. However, if such an additional population were introduced into the 
modelling, in principle, they should be bluer objects rather than 
ellipticals; otherwise, such models would overpredict the number counts for 
ellipticals. But if the additional population were blue objects, even 
though they would not make the predicted color distribution even worse, the 
predictions can not be ameliorated beyond $B-K=5.5$ in the color 
distribution, and especially, the red-end humps beyond $B-K=8.0$ can not be 
depressed. From the investigation of the color distribution, we arrive at 
the conclusion that there is no room for an additional galaxy population to 
be introduced into the models without first revising the traditional 
formation and evolution scenario for ellipticals.

We have considered a simple number-evolution model phenomenologically to 
show that number evolution could be the right answer to the question 
regarding the formation and evolution of elliptical galaxies. In 
particular, if the cosmic geometry is open or $\Lambda$-dominated, our work 
supports the idea that ellipticals formed by mergers of spiral galaxies, 
which is compatible with the prediction of hierarchical theory for 
formation and evolution of galaxies. In a forthcoming work, we will 
constructed a unifying number-evolution model to explain the observations 
concerning the faint galaxies at high redshifts.

\acknowledgements

PH gratefully acknowledge Mrs. G. Kauffmann for her kindly help, and Mr. S. 
Charlot for providing us with their latest spectral synthesis models. We 
thank Mr. Jon Loveday for his careful reading of the manuscript, and for 
his constructive comments and suggestions to improve this paper. We also 
thank The State Key Laboratory of Science and Engineering Computing (LSEC) 
of Academia Sinica for providing us with computer supports. This work is in 
part supported by the National Natural Science Foundation of China.

\newpage

\begin{deluxetable}{lccccc}
\tablecolumns{6}
\tablewidth{30pc}
\tablecaption{Observed Colors of galaxies}
\tablehead{
\colhead{Type} & \colhead{$U-B$\tablenotemark{a}} & 
\colhead{$B-V$\tablenotemark{a}} &
\colhead{$B-R$\tablenotemark{b}} & \colhead{$b_j-I$\tablenotemark{c}} &
\colhead{$b_j-K$\tablenotemark{b}}}
\startdata
E/S0   &   0.43  &  0.95   &   1.83  &  2.39  &  4.16 \nl
Sab    &   0.19  &  0.79   &   1.65  &  2.15  &  3.88 \nl
Sbc    &   0.07  &  0.68   &   1.30  &  1.78  &  3.64 \nl
Scd    &  -0.04  &  0.60   &   1.16  &  1.56  &  3.23 \nl
Sdm    &  -0.17  &  0.47   &   1.05  &  1.27  &  2.80
\enddata
\tablenotetext{a}{Mixed from Fukugita et al. (1995);}
\tablenotetext{b}{Mixed from Yoshii et al. (1988);}
\tablenotetext{c}{Mixed from Yoshii et al. (1988, modelled) and Lidman et al. 
(1996).}
\end{deluxetable}

\begin{deluxetable}{ccccccccc}
\tablecolumns{9}
\tablewidth{0pc}
\tablecaption{Extinction magnitudes for $B$, $I$, and $K$ bands at different 
redshifts}
\tablehead{
 & \multicolumn{2}{c}{$B$}&  &\multicolumn{2}{c}{$I$} &   & 
\multicolumn{2}{c}{$K$} \\
\cline{2-3} \cline{5-6} \cline{8-9} \\
\colhead{$z$} & \colhead{E/S0\tablenotemark{a}} & 
\colhead{Others\tablenotemark{a}}
&  & \colhead{E/S0\tablenotemark{a}}&\colhead{Others\tablenotemark{a}}& 
 &\colhead{E/S0\tablenotemark{a}} & \colhead{Others\tablenotemark{a}}} 
\startdata
0   &   0.05 &   0.10  &     &   0.02  &  0.03  &    &   0.002   &   0.005  \nl
1   &   0.20 &   0.36  &     &   0.07  &  0.13  &    &   0.01    &   0.02  \nl
2   &   0.42 &   0.64  &     &   0.15  &  0.27  &    &   0.02    &   0.04  \nl
3   &   0.65 &   0.83  &     &   0.25  &  0.44  &    &   0.04    &   0.07  \nl
4   &   0.86 &   0.92  &     &   0.37  &  0.60  &    &   0.06    &   0.11  \nl
5   &   1.02 &   0.97  &     &   0.50  &  0.73  &    &   0.08    &   0.15
\enddata
\tablenotetext{a}{The optical depth $\tau^*_0$=0.10, geometrical parameter 
                  $\zeta$=0.50 for ellipticals; $\tau^*_0$=0.20, $\zeta$=0.25
                  for the other types.}
\end{deluxetable}

\begin{deluxetable}{lccccccccc}
\tablecolumns{10}
\tablewidth{35pc}
\tablecaption{Model parameters and predicted colors}
\tablehead{
      &       &   &   & Scenario A        &       &        &       &      & \\
Type & $\tau_e$\tablenotemark{a} & $U-B$ & $B-V$ & $B-R$ & $b_j-I$ & $b_j-K$ &
$\alpha$\tablenotemark{b}&$\phi^*$\tablenotemark{b}&$M^*_B$\tablenotemark{b}}
\startdata
E/S0  &  0.2  & 0.65  &  0.92   &   1.69  &  2.30  &  4.15 & -0.70 & 9.68 & 
-21.00 \nl
Sab   &  2.3  & 0.46  &  0.83   &   1.55  &  2.15  &  3.89 & -1.25 & 3.85 & 
-20.80 \nl
Sbc   &  4.6  & 0.11  &  0.61   &   1.24  &  1.81  &  3.45 & -1.40 & 6.25 & 
-20.80 \nl
Scd   &  9.5  & -0.05 &  0.49   &   1.04  &  1.57  &  3.14 & -1.45 & 5.75 & 
-20.80 \nl
Sdm  &$\infty$& -0.15 &  0.39   &   0.88  &  1.37  &  2.87 & -1.50 & 2.85 & 
-20.70 \nl
\cline{1-10} \\ 
      &       &   &   & Scenario B        &        &       &       &      & \\
Type & $\tau_e$\tablenotemark{a} & $U-B$ & $B-V$ & $B-R$ & $b_j-I$ & $b_j-K$ &
$\alpha$\tablenotemark{b}&$\phi^*$\tablenotemark{b}&$M^*_B$\tablenotemark{b}
\vspace{3pt}\nl
\cline{1-10}\vspace{-6pt}
      &       &       &         &         &        &       &       &      &      
  \nl
E/S0  &  0    & 0.71  &  0.95   &   1.74  &  2.36  &  4.25 & -0.70 & 9.68 & 
-21.00 \nl
Sab   &  3.3  & 0.43  &  0.82   &   1.55  &  2.17  &  3.93 & -0.95 & 4.05 & 
-20.80 \nl
Sbc   &  6.3  & 0.11  &  0.62   &   1.26  &  1.84  &  3.51 & -1.10 & 6.05 & 
-20.80 \nl
Scd   & 13.5  & -0.04 &  0.49   &   1.06  &  1.60  &  3.19 & -1.50 & 5.75 & 
-20.80 \nl
Sdm  &$\infty$& -0.13 &  0.41   &   0.92  &  1.42  &  2.95 & -1.60 & 2.85 & 
-20.70 \nl
\cline{1-10} \\ 
      &       &   &   & Scenario C        &        &       &       &      & \\
Type & $\tau_e$\tablenotemark{a} & $U-B$ & $B-V$ & $B-R$ & $b_j-I$ & $b_j-K$ &
$\alpha$\tablenotemark{b}&$\phi^*$\tablenotemark{[b,c]}&$M^*_B$\tablenotemark{[b,c]}
\vspace{3pt}\nl
\cline{1-10}\vspace{-6pt}
      &       &       &         &         &        &       &       &      &      
  \nl
E/S0  &  0.05 & 0.72  &  0.95   &   1.74  &  2.37  &  4.26 & -0.70 & 9.68 & 
-21.00 \nl
Sab   &  3.5  & 0.40  &  0.81   &   1.54  &  2.15  &  3.91 & -0.95 & 3.85 & 
-20.80 \nl
Sbc   &  7.0  & 0.08  &  0.60   &   1.23  &  1.81  &  3.47 & -1.05 & 6.25 & 
-20.80 \nl
Scd   & 15.0  & -0.05 &  0.49   &   1.06  &  1.59  &  3.18 & -1.45 & 5.75 & 
-20.80 \nl
Sdm  &$\infty$& -0.13 &  0.41   &   0.92  &  1.43  &  2.96 & -1.55 & 2.85 & 
-20.70
\enddata
\tablenotetext{a}{$\tau_e$ is the timescale of SFR, measured in Gyr;}
\tablenotetext{b}{$\alpha$, $\phi^*$, and $M^*_B$ are parameters for LFs, 
$\phi^*$ in units of 10$^{-4}$Mpc$^{-3}$;}
\tablenotetext{c}{The Hubble constant has been scaled to $h=0.5$ for Scenario C.}
\end{deluxetable}

\begin{deluxetable}{cccc}
\tablecolumns{4}
\tablewidth{30pc}
\tablecaption{Parameters of Merger Model.}
\tablehead{
\colhead{Model parameter} & \colhead{Scenario A} & \colhead{Scenario B} &
\colhead{Scenario C}}
\startdata
$z_t$    &   5.0  &  1.0  &  1.0    \nl
$Q_{n1}$ &   0.05 &  0.05 &  0.1    \nl
$Q_{n2}$ &   ...  &  2.0  &  2.0    \nl
$Q_m$    &   0.3  &  0.2  &  0.1
\enddata
\end{deluxetable}

\begin{figure*}[htb]
\centerline{\psfig{figure=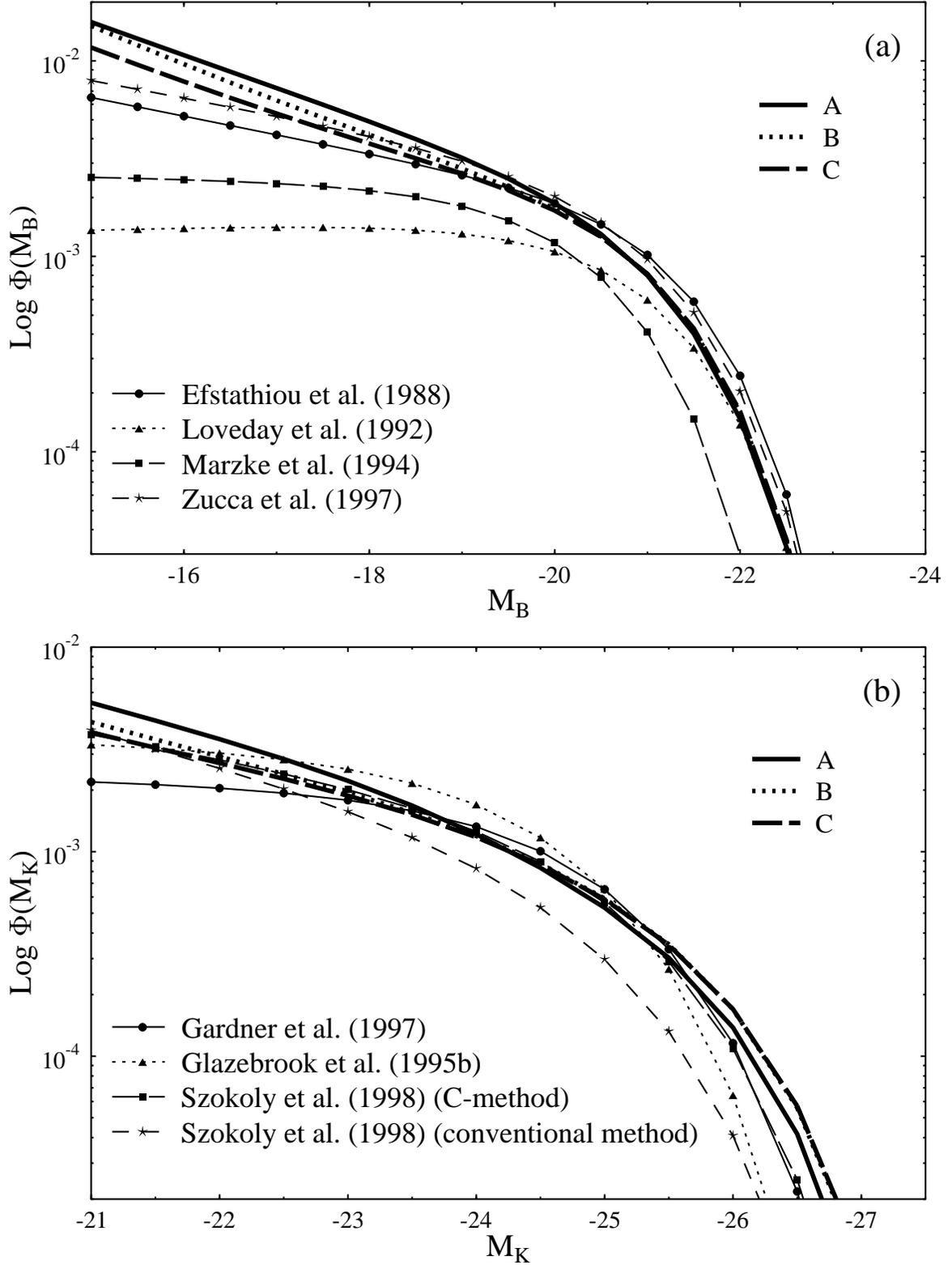,angle=0,width=16.0cm}}
\figcaption{The present-day luminosity functions of galaxies. The models are indicated 
by thick lines, whose values are corresponding to those in Table 3, and the capital 
letters A, B, and C refer to Scenario A, B, and C respectively. The observational data 
are shown by thin lines decorated with various markers, and the sources of these 
observations are also indicated in the figure. Notice that the characteristic density 
$\phi^*$ and mixing ratio between different galaxies of Efstathiou et al. (1988) LF are 
taken from Pozzetti et al. (1996). Panels (a) and (b) are for $B$- and $K$-bands 
respectively. The Hubble constant has been scaled to $h=0.5$ for Scenario C.
}
\end{figure*}

\begin{figure*}[htb]
\centerline{\psfig{figure=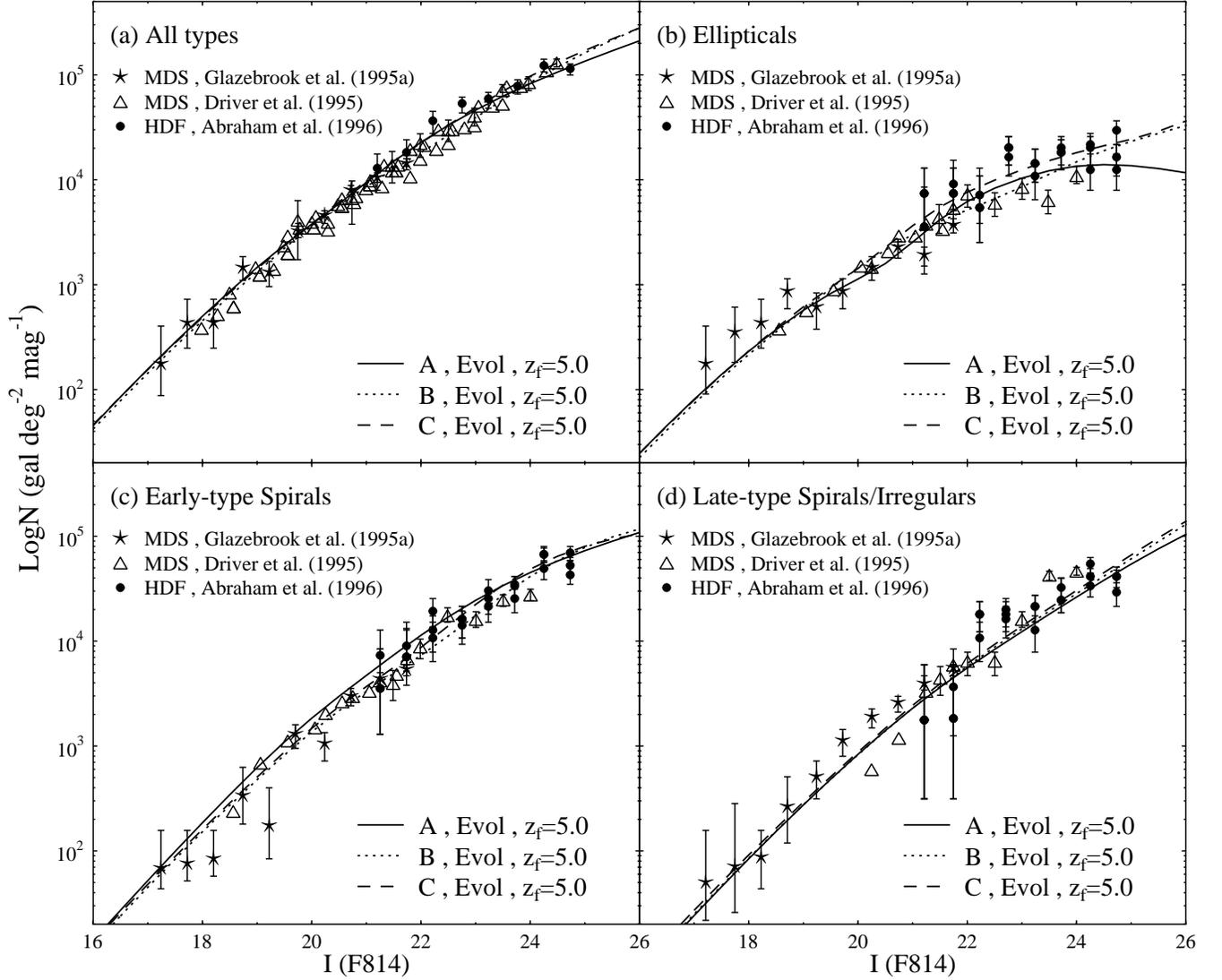,angle=0,width=18cm}}
\figcaption{Differential number counts as a function of apparent magnitude in
$I_{814}$-band. The sources of observational data are exhibited in the figure 
and these data are indicated by symbols. Predictions of models are shown by lines. 
Panels (b), (c), and (d) are for ellipticals, early-type spirals, and late-type 
spirals/irregulars respectively, and (a) for the overall population. Here and 
hereafter, capital letters A, B, and C represent Scenario A, B, and C, 
respectively.
}
\end{figure*}

\begin{figure*}[htb]
\centerline{\psfig{figure=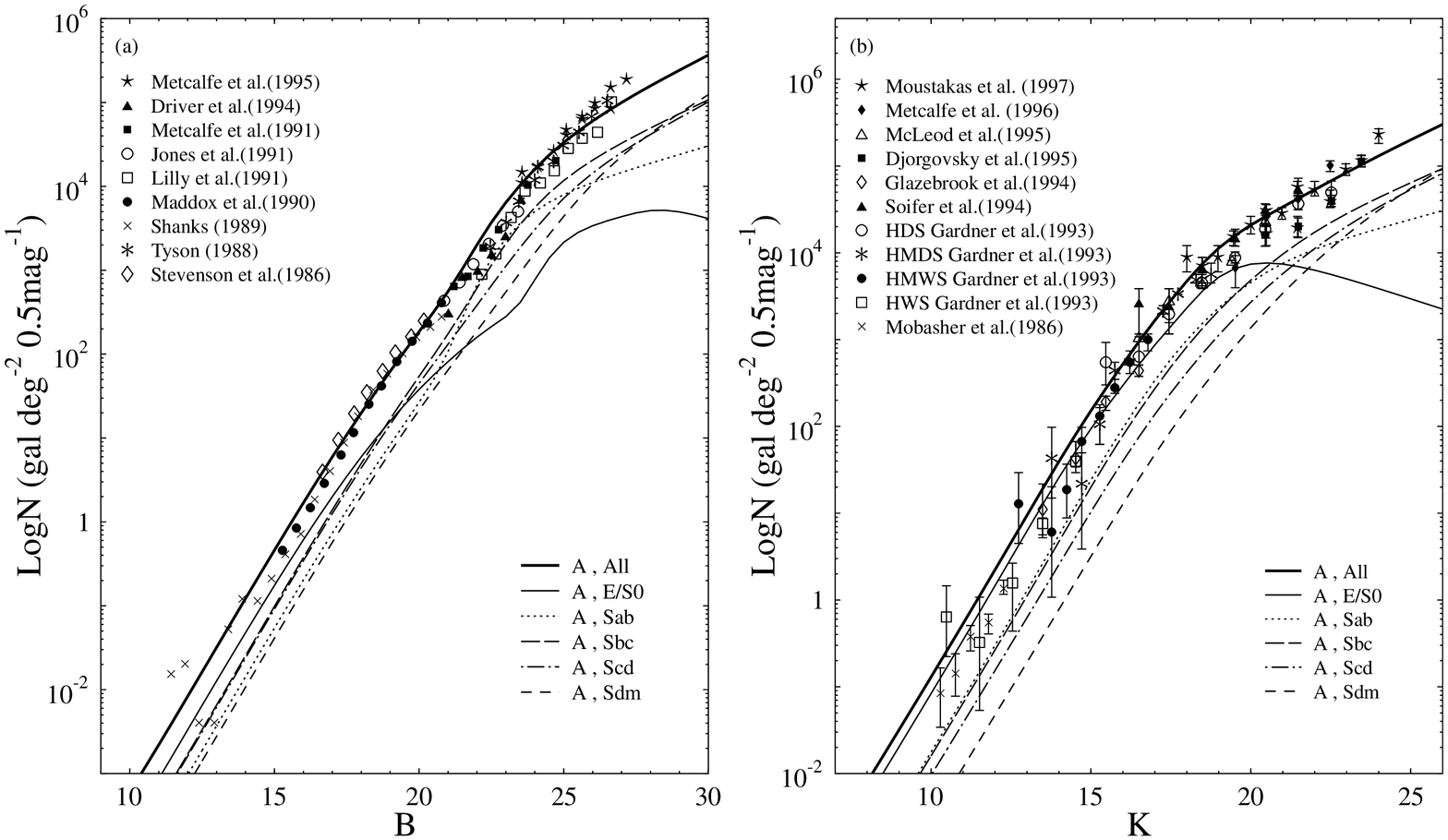,angle=0,width=18.5cm}}
\end{figure*}

\begin{figure*}[htb]
\centerline{\psfig{figure=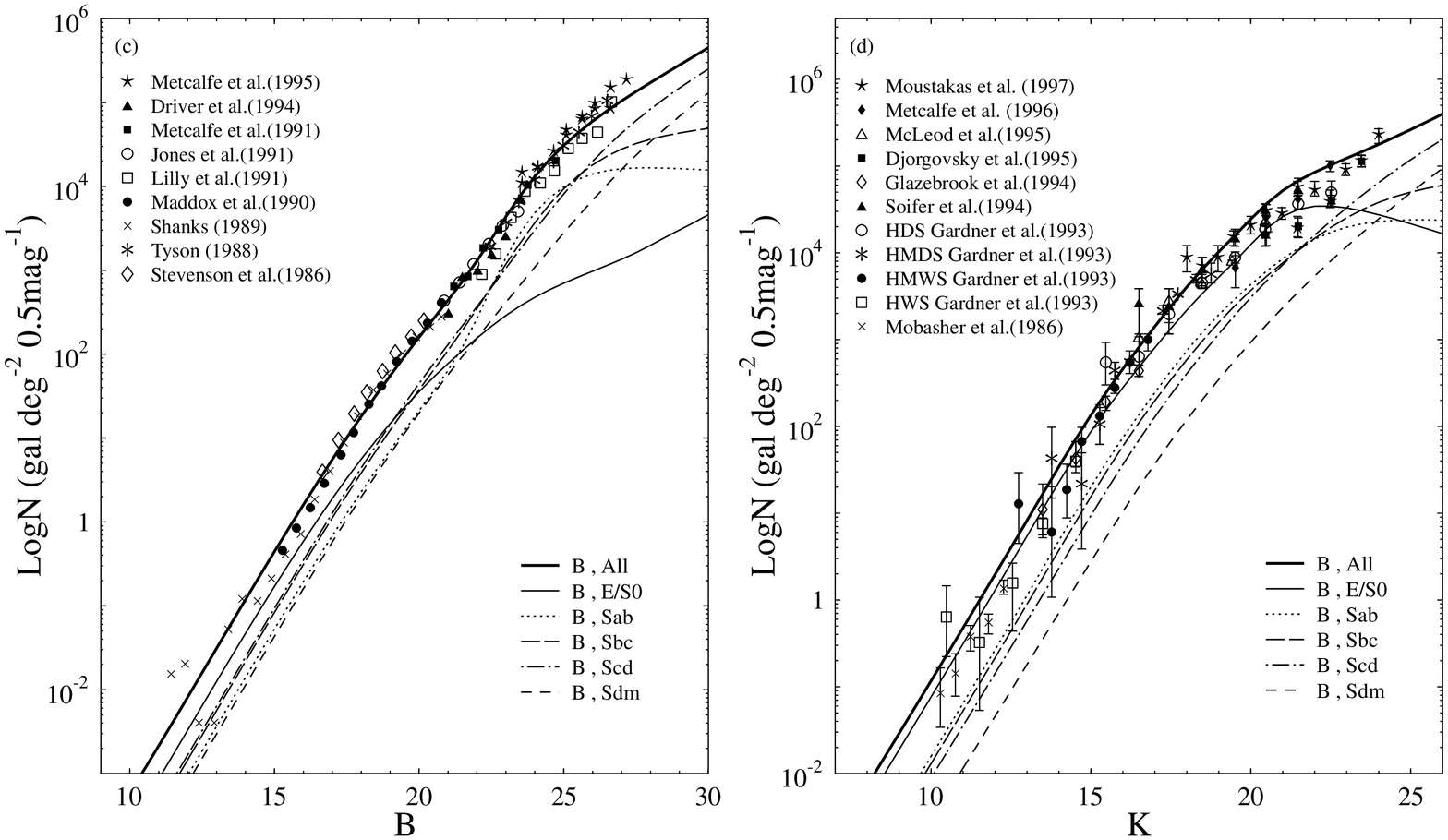,angle=0,width=18.5cm}}
\end{figure*}

\begin{figure*}[htb]
\centerline{\psfig{figure=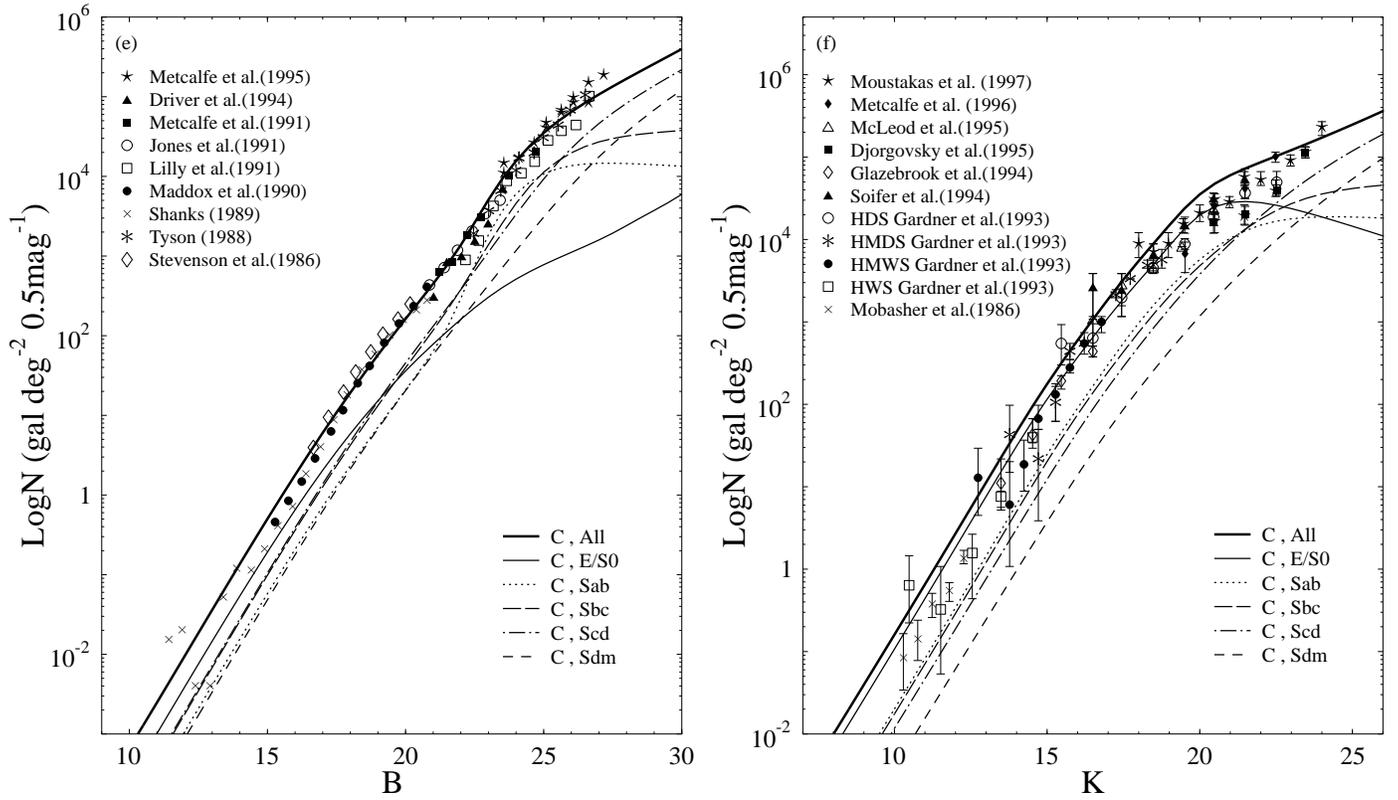,angle=0,width=18.5cm}}
\figcaption{Differential number counts as a function of apparent magnitude in 
$B$- (panel a, c, and e) and $K$- (panel b, d, and f) bands. Panels (a) and (b) 
are for Scenario A, (c) and (d) for Scenario B, and (e) and (f) for Scenario C, 
respectively. The sources of observational data are exhibited in the figure. 
Predictions of models are shown by lines.
}
\end{figure*}

\begin{figure*}[htb]
\centerline{\psfig{figure=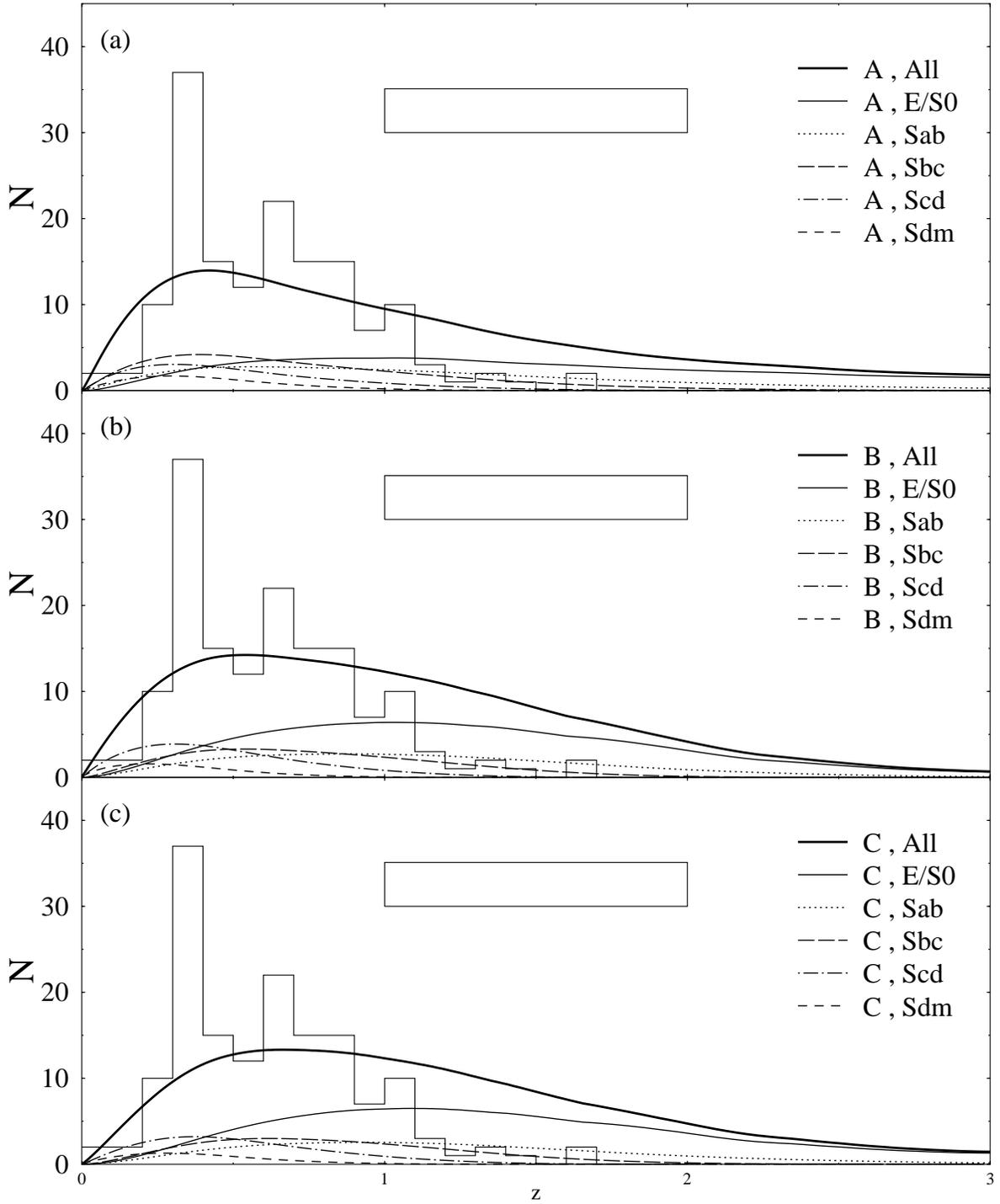,angle=0,width=16cm}}
\figcaption{Redshift distribution for galaxies limited to $K<20$. The 
observational data are derived from Cowie et al. (1996), and are shown by the 
solid histograms. Predictions of models are shown by lines. The model 
predictions have been normalized to the total number of both $z$-identified and 
$z$-unidentified objects ($N_{id}+N_{no-id}$). The area of the rectangle in each 
panel equals the number of unidentified objects ($N_{no-id}$). Panels (a), 
(b), and (c) are for Scenario A, B, and C respectively.
}
\end{figure*}

\begin{figure*}[htb]
\centerline{\psfig{figure=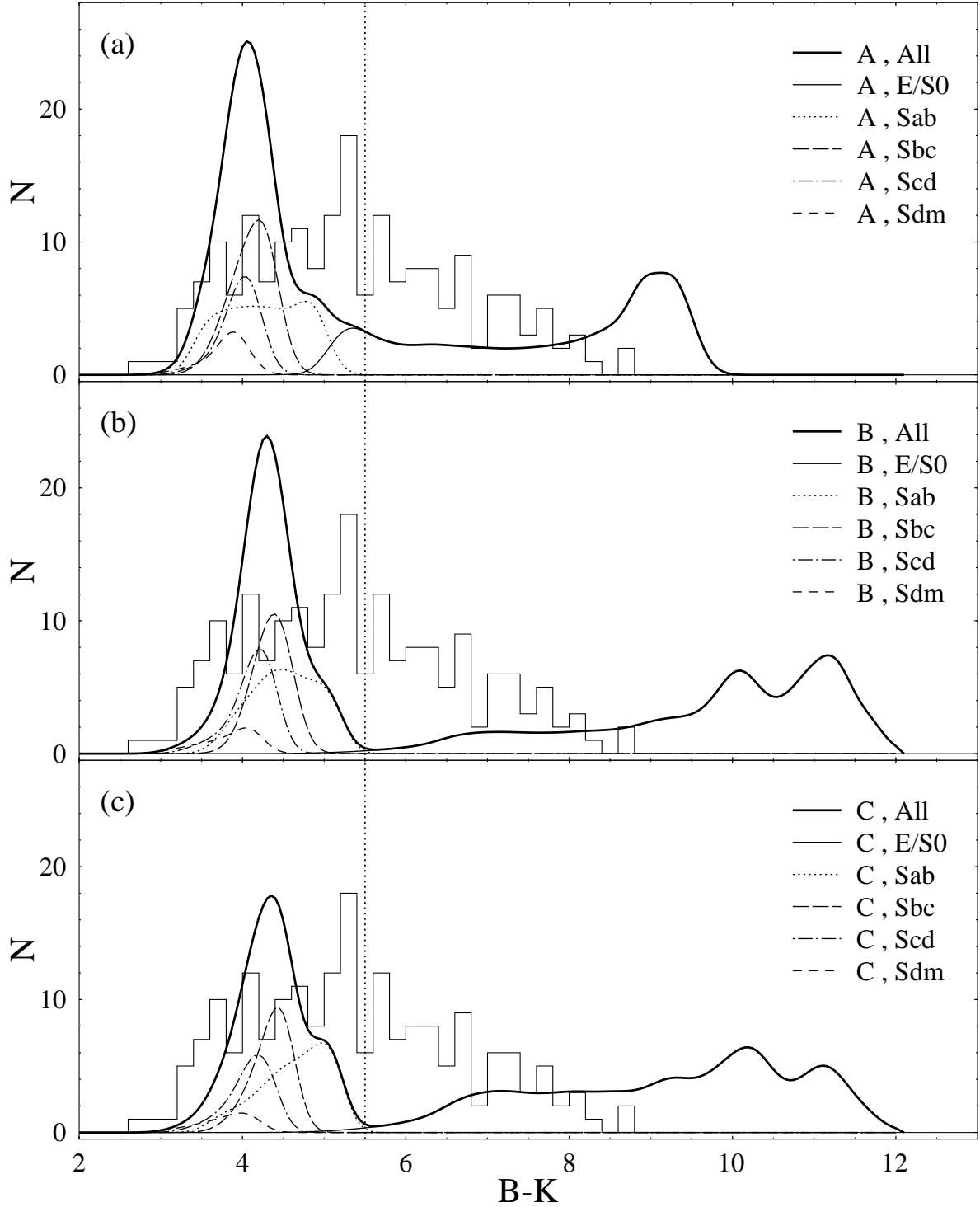,angle=0,width=16.0cm}}
\figcaption{$B-K$ color distribution for $17<K<20$. The solid histogram shows 
the observed distribution derived from Cowie et al. (1996). The predictions are 
shown by various lines. Panels (a), (b), and (c) are for Scenario A, B, and C, 
respectively. To aid the eye, we draw a vertical dotted line at $B-K=5.5$ for 
the reason explained in the text. The model predictions have been normalized to 
the total number of objects enclosed by the histogram.
}
\end{figure*}

\begin{figure*}[htb]
\centerline{\psfig{figure=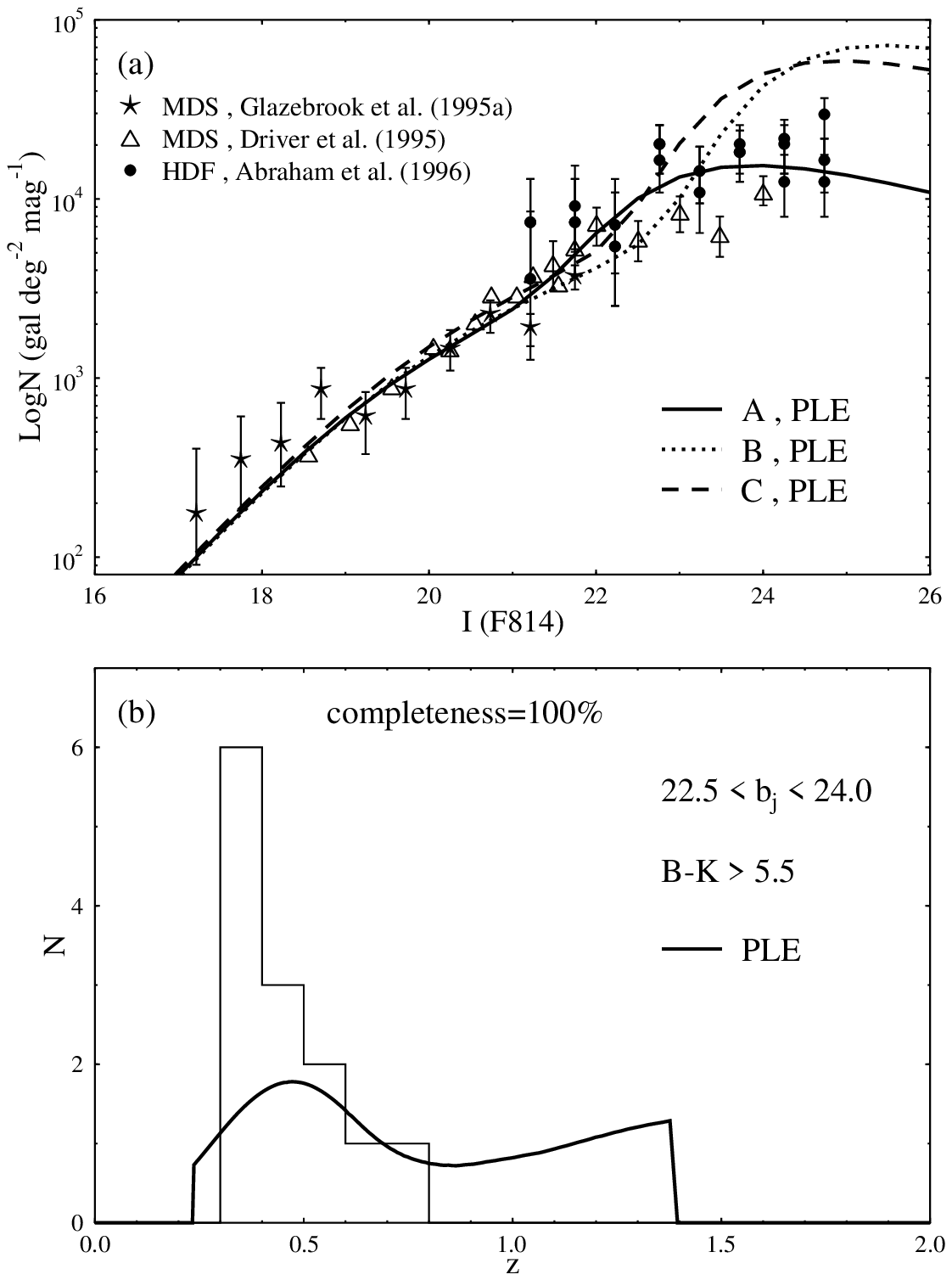,angle=0,width=16.0cm}}
\figcaption{(a) Differential number counts as a function of apparent 
magnitude in $I_{814}$-band. The sources of observational data are exhibited 
in the figure and these data are indicated by symbols. Predictions of pure luminosity 
evolution (PLE) models 
are shown by lines. (b) Redshift distribution for galaxies limited to 
$22.5<b_{j}<24.0$ and with the constraint of $B-K>5.5$. The observational 
data are derived from Cowie et al. (1996), which are shown by the solid 
histogram. The model prediction is shown by the solid line, and it has been 
normalized to the total number of the objects enclosed by the histogram.
}
\end{figure*}

\begin{figure*}[htb]
\centerline{\psfig{figure=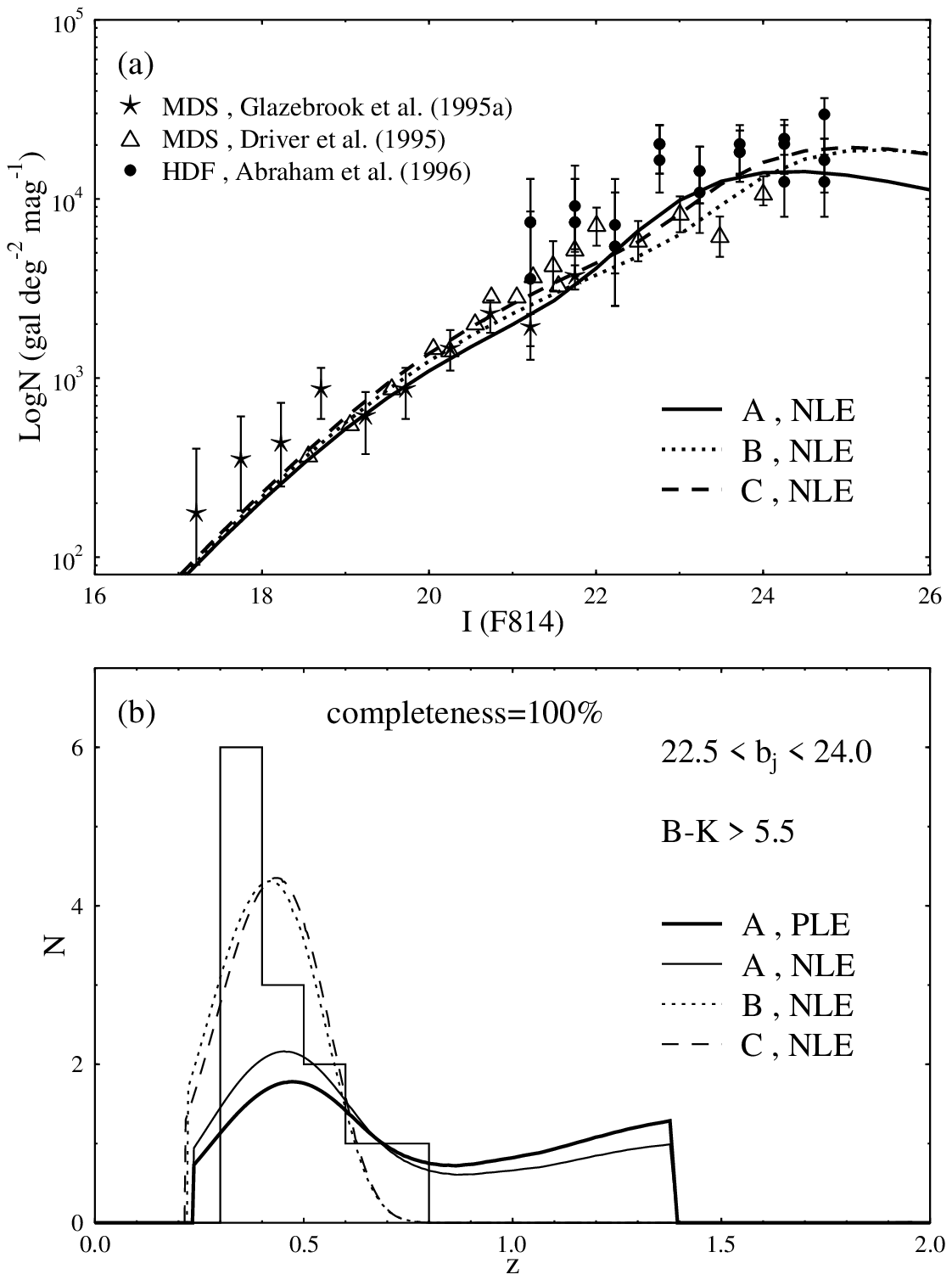,angle=0,width=16.0cm}}
\figcaption{(a) Differential number counts as a function of apparent 
magnitude in $I_{814}$-band. Predictions of number-luminosity evolution (NLE) models 
are shown by lines. (b) Redshift distribution for galaxies limited to $22.5<b_{j}<24.0$ 
and with the constraint of $B-K>5.5$. The model predictions are shown by lines, and 
they have been normalized to the total number of the objects enclosed by the histogram.
}
\end{figure*}

\begin{figure*}[htb]
\centerline{\psfig{figure=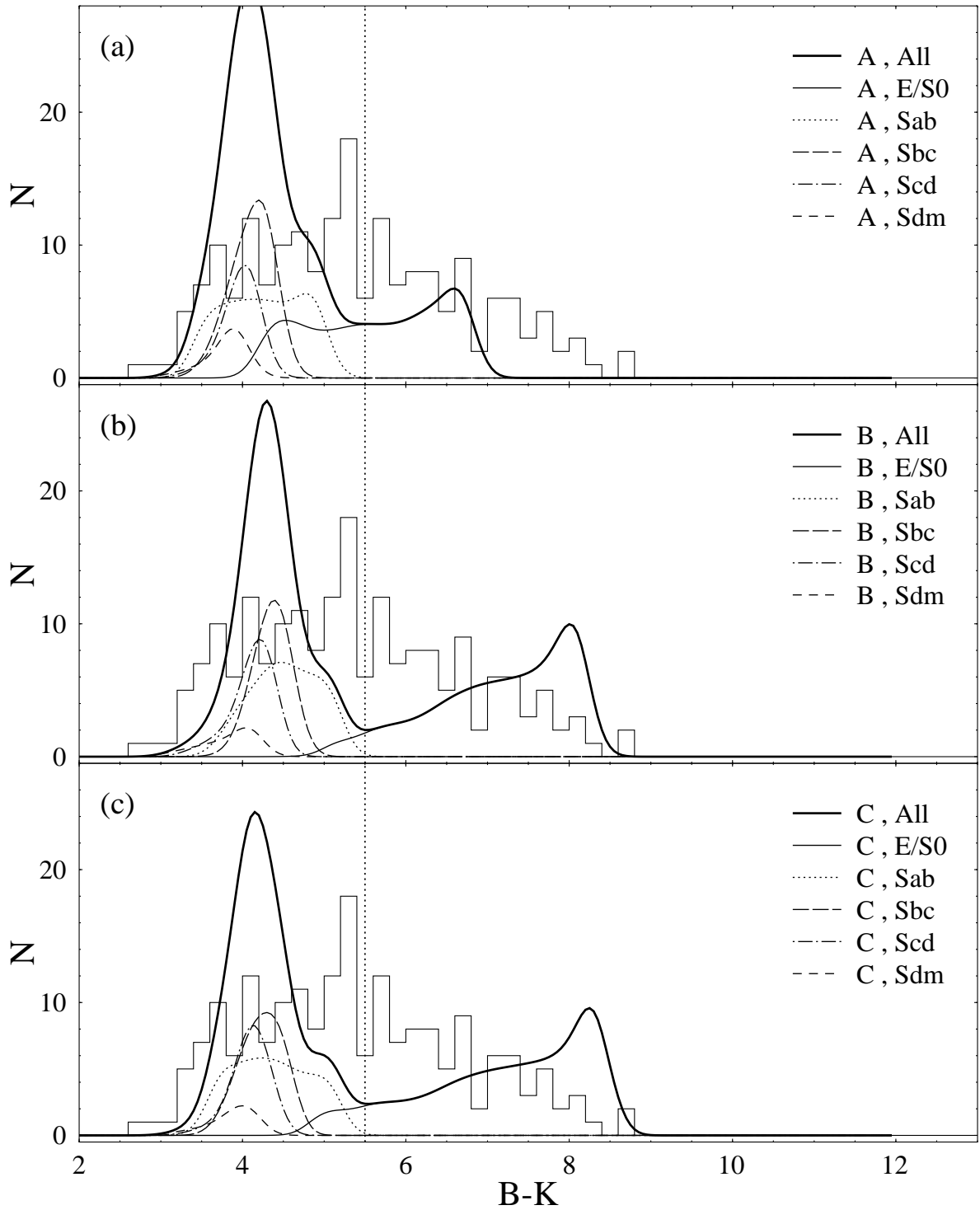,angle=0,width=16.0cm}}
\figcaption{The same as Figure 5, but the predictions are made by the NLE assumption 
for ellipticals, described by Eq. [2] and [3].
}
\end{figure*}

\end{document}